\documentclass[9pt,conference]{IEEEtran}
\usepackage{bm}
\usepackage{tabularx}
\usepackage{tablefootnote}
\usepackage{threeparttable}
\usepackage{colortbl}
\usepackage{comment}
\usepackage{here}
\usepackage{algorithm, algorithmic}
\usepackage{svg}
\usepackage{siunitx}
\usepackage{hyperref}
\usepackage{amsfonts}
\usepackage{amsmath}
\usepackage{booktabs}
\usepackage{cleveref}

\usepackage{xcolor}

\title{Schr\"odinger Bridge Consistency Trajectory Models\\ for Speech Enhancement}

\author{Shuichiro Nishigori$^{1}$, Koichi Saito$^{2}$, Naoki Murata$^{3}$, Masato Hirano$^{1}$, Shusuke Takahashi$^{1}$, and Yuki Mitsufuji$^{1,2}$\\
$^{1}$ Sony Group Corporation, Tokyo, Japan \; $^{2}$ Sony AI, New York, USA \; $^{3}$ Sony AI, Tokyo, Japan}

\begin{document}

\maketitle

\begin{abstract}
Speech enhancement (SE) using diffusion models is a promising technology for improving speech quality in noisy speech data. Recently, the Schr\"odinger bridge (SB) has been utilized in diffusion-based SE to resolve the mismatch between the endpoint of the forward process and the starting point of the reverse process. However, the SB inference remains slow owing to the need for a large number of function evaluations (NFE) to achieve high-quality results. Although consistency models have been proposed to accelerate inference, by employing consistency training via distillation from pretrained models in the field of image generation, they struggle to improve generation quality as the number of steps increases. Consistency trajectory models (CTMs) have been introduced to overcome this limitation, achieving faster inference while preserving a favorable trade-off between quality and speed. In particular, SoundCTM applies CTM techniques to sound generation. The SB addresses the aforementioned mismatch and CTM improves inference speed while maintaining a favorable trade-off between quality and speed. However, no existing method simultaneously resolves both issues to the best of our knowledge. Hence, in this study, we propose Schr\"odinger bridge consistency trajectory models (SBCTM), which apply CTM techniques to the SB for SE. In addition, we introduce a novel auxiliary loss, incorporating perceptual loss, into the original CTM training framework. Consequently, SBCTM achieves an approximately $16\times$ improvement in real-time factor compared with the conventional SB for SE. Furthermore, the favorable trade-off between quality and speed in SBCTM enables time-efficient inference by limiting multi-step refinement to cases in which one-step inference is insufficient.
SBCTM enables more practical and efficient speech enhancement by providing fast inference and a flexible mechanism for further quality improvement.
Our codes, pretrained models, and audio samples are available at
\href{https://github.com/sony/sbctm/}{https://github.com/sony/sbctm/}.
\end{abstract}

\section{INTRODUCTION} \label{sec:intro}

Speech signals have recently become easily recorded using smartphones and other recording tools. However, recorded speech signals frequently contain background noise, which negatively affects speech intelligibility and sound quality. The primary purpose of speech enhancement (SE) is to clean noisy speech signals. Deep neural network-based SE methods generally use deep neural network models to learn speech signal features from training data. While both discriminative methods~\cite{demucs,mpsenet} and generative models~\cite{cmgan} are employed, generative models in particular contribute to improving the performance by modeling the distribution of clean speech signals in noisy environments.
Recently, diffusion models (DMs)~\cite{diffusion} have been used for image generation and SE~\cite{cdpmse,sgmse0,sgmse,storm,rsde,review}. DMs train on multiple noise levels, learning robust representations that generalize across diverse conditions. This regularization effect makes them well-suited for tasks like SE. Score-based generative model for speech enhancement (SGMSE+)~\cite{sgmse} illustrates that DMs have the advantage of a better generalization performance than other generative models. However, recent SE methods, particularly diffusion-based methods, have some limitations: (1) prior mismatch and (2) low inference speed. 

Diffusion-based SE is known to suffer from a mismatch between the endpoint of the forward process and the starting point of the reverse process (\textit{prior mismatch} problem)~\cite{rsde}. Several approaches have been proposed to address this issue, including diffusion denoising bridge models (DDBMs)~\cite{ddbm} and the Schr\"odinger bridge (SB)~\cite{sb}. In particular, the SB, a framework for probabilistic transportation that seeks an optimal way to transport one probability distribution to another, has been used in diffusion-based SE~\cite{sb-se}. The SB can start the reverse process directly from the noisy speech signal, thereby avoiding the prior mismatch problem. Furthermore, SB-PESQ~\cite{sgmse-sb,sb-fs48k} achieves improved perceptual quality by incorporating a perceptual loss during training. While the method is originally proposed in~\cite{sgmse-sb}, the name 'SB-PESQ' is later introduced in~\cite{sb-fs48k}.
DMs also exhibit low inference speed because a sequential reverse diffusion process is necessary to achieve sufficient quality. Consistency models (CMs)~\cite{cm} present a distillation training framework for improving the quality of one-step image generation to accelerate the inference speed of DMs. For image-to-image tasks, consistency diffusion bridge models (CDBMs)~\cite{cdbm} apply the CM training framework to the DDBM to sample images $4\times$ to $50\times$ faster than the original DDBM. SE-bridge~\cite{seb} utilizes the CM training framework for SE to accelerate the inference speed by $15\times$ compared with the original SGMSE+ as their primary baseline. Although consistency training can accelerate the inference speed, it does not improve the generation quality when the number of steps increases~\cite{ctm}. As a solution to this problem, consistency trajectory models (CTMs)~\cite{ctm} propose a model that not only enables fast image generation but also achieves a favorable trade-off between quality and speed. Therefore, multi-step refinement is required only when one-step inference is insufficient. This selective refinement strategy significantly reduces the total inference time compared with always performing multi-step inference. Sound consistency trajectory models (SoundCTM)~\cite{sctm} utilize the CTM technique to text-to-sound (T2S) generation to achieve both one-step high quality sound generation and multi-step higher-quality sound generation, demonstrating that the CTM's technique can be applied not only to image generation but also to sound generation.

The SB addresses prior mismatch and CTM improves inference speed while maintaining a favorable trade-off between quality and speed. However, to the best of our knowledge, no method exists that resolves both of them thus far.

In this study, inspired by the achievements of SoundCTM as an application of the CTM to the field of sound generation, we propose Schr\"odinger bridge consistency trajectory models (SBCTM), introducing the CTM's training framework into the SB for SE. This contributes to the SE field by enabling controllability of both quality and speed. Our contributions can be summarized as follows:
\begin{itemize}
    \item We introduce the CTM's technique to the SB for SE, achieving approximately $16\times$ improvement in the real-time factor (RTF) compared with the original SB while maintaining a favorable trade-off between quality and speed.
    \item We introduce a novel auxiliary loss including perceptual loss to the training framework of the CTM to achieve high-quality SE.
\end{itemize}
The remainder of this paper is organized as follows. Section~\ref{sec:pre} reviews the necessary preliminaries, including the SB and CTM. Section~\ref{sec:methods} introduces our proposed model, SBCTM. Section~\ref{sec:exp} describes the experimental setup used to evaluate the proposed method. Section~\ref{sec:res} presents and discusses the experimental results. Finally, Section~\ref{sec:con} concludes the paper.

\section{PRELIMINARIES} \label{sec:pre}

\subsection{Schr\"odinger bridge for speech enhancement}

As discussed in Section~\ref{sec:intro}, standard DMs for SE often suffer from a distribution mismatch. In these models, the forward process uses an artificial prior, such as white Gaussian noise, as the terminal distribution, which differs significantly from real noisy speech~\cite{sgmse}. This mismatch degrades the performance and increases the inference cost.

The SB framework~\cite{sb} provides a solution by seeking a stochastic process $\mathbb{Q}$, that is, a continuous-time evolution of probability distributions that transports one realistic distribution to another. Specifically, given the clean speech distribution $p_0$ and noisy speech distribution $p_1$, SB finds a process that minimizes the KL divergence from a simple reference process $\mathbb{P}$ while satisfying the marginal constraints, implying that the initial and terminal distributions must match $p_0$ and $p_1$, respectively:
\begin{equation}
    \min_{\mathbb{Q}} \mathrm{KL}(\mathbb{Q} \,\Vert\, \mathbb{P}) \quad \text{subject to} \quad \mathbb{Q}_{t=0} = p_0, \quad \mathbb{Q}_{t=1} = p_1.
\end{equation}
In the context of DMs, the reference process $\mathbb{P}$ is typically selected as a simple diffusion process, such as Brownian motion. Standard score-based DMs aim to approximate the reverse-time dynamics of $\mathbb{P}$ perturbed by data-specific noise and effectively solve a related but different problem: they bridge an artificial prior distribution (e.g., standard Gaussian noise) to a clean data distribution, rather than connecting two realistic distributions.

This distinction motivates the use of the SB formulation for SE tasks, where both endpoints---clean speech and noisy speech---are real and observable. By directly modeling the stochastic transport between these two realistic distributions, we can better align the forward and reverse processes without relying on unrealistic assumptions regarding the data.

This formulation enables the construction of a realistic transformation path between clean and noisy speech without introducing artificial priors. Motivated by this property, recent studies applied the SB framework to SE by designing DMs that operate directly between clean and noisy data. Rather than training the model to predict the score function of an artificial noise distribution, it is possible to predict clean speech samples directly through data prediction loss~\cite{sb-se, sgmse-sb}
\begin{equation}
    \mathcal{L}_{\text{data}} = \mathbb{E}_{t \sim \mathcal{U}[0, 1], x_0 \sim p_0, x_1 \sim p_1}[\| x_0 - x_\theta(x_t, x_1, t) \|_2^2],
\end{equation}
where $x_0$ and $x_1$ are sampled from the start and end distributions $p_0$ and $p_1$, respectively. $t$ represents the time index. The noisy sample $x_t$ is generated from the clean sample $x_0$ through a forward diffusion process. By minimizing this loss function, the model $x_\theta$ learns to predict the clean speech $x_0$ from the noisy speech $x_t$. For notational simplicity, we omit the expectation operator $\mathbb{E}$ in subsequent equations.

By starting the reverse process from actual noisy speech, instead of from an artificial prior, this approach improves both the quality and efficiency of the enhancement. Furthermore, it enables the incorporation of auxiliary objectives, such as time-domain reconstruction~\cite{sb-se, sgmse-sb} or perceptual quality metrics~\cite{sgmse-sb}, to further enhance the performance.

\subsection{Consistency Trajectory Models and SoundCTM}

CTM~\cite{ctm} predicts points from any given point on the probability flow ordinary differential equation (PF ODE) trajectory~\cite{ncsnpp} expressed as the deterministic trajectory of the reverse diffusion process. $G(x_t, t, s)$ is defined as the solution of the PF ODE from the start time $t$ to the end time $s \leq t$ and $G$ is estimated by a neural network model $G_\theta$. For training $G_\theta$, the CTM loss~\cite{ctm} is calculated by comparing $x_\mathrm{tgt} = G_\mathrm{sg(\theta)}(G_\mathrm{sg(\theta)}(\mathrm{Solver}(x_t, t, u; \phi), u, s), s, 0)$ as the prediction from teacher $\phi$ with $x_\mathrm{est} = G_\mathrm{sg(\theta)}(G_\theta(x_t, t, s), s, 0)$ as the prediction from student $\theta$, where $\mathrm{Solver}(x_t, t, s;\phi)$ denotes the pretrained PF ODE sampler, $u \in [s, t)$ denotes the amount of distillation from the teacher, and
$\mathrm{sg}(\cdot)$ denotes the exponential moving average stop gradient $\mathrm{sg(\theta)} \leftarrow \mathrm{stopgrad}(\mu \, \mathrm{sg(\theta)} + (1 - \mu)\theta)$. Furthermore, the denoising score matching (DSM) loss~\cite{ctm} and the GAN loss~\cite{ganloss} have been introduced as auxiliary losses. By following SoundCTM~\cite{sctm} and eliminating the GAN loss owing to the sensitivity of the discriminator selection, the losses are defined as follows:
\vspace{-3pt}
\begin{align}
    \mathcal{L}_\mathrm{CTM} &= d_\text{feat}(x_\mathrm{tgt}(x_t, t, u, s), x_\mathrm{est}(x_t, t, s)), \label{eq:org_ctmloss} \\
    \mathcal{L}_\mathrm{DSM} &= \|x_0 - g_\theta(x_t, t, t)\|_2^2, \label{eq:org_dsmloss}
\end{align}
where $d_\mathrm{feat}$ is a function that measures feature distance. The objectives are defined as follows:
\begin{equation}
    \mathcal{L} = \mathcal{L}_\mathrm{CTM} + \lambda_\mathrm{DSM}\mathcal{L}_\mathrm{DSM}, \label{eq:orgloss}
\end{equation}
where $\lambda_\mathrm{DSM}$ denotes the weight used to balance the two terms.

SoundCTM~\cite{sctm} achieved high-quality sound generation by addressing the limitations of the CTM's training framework. It proposes a domain-agnostic feature distance, methods for distilling classifier-free guided trajectories, and techniques for text-conditional and unconditional jumps in sampling.

\section{METHODS} \label{sec:methods}

\subsection{Model training}

\begin{figure}[H]
    \centering
    \includegraphics[width=0.8\linewidth]{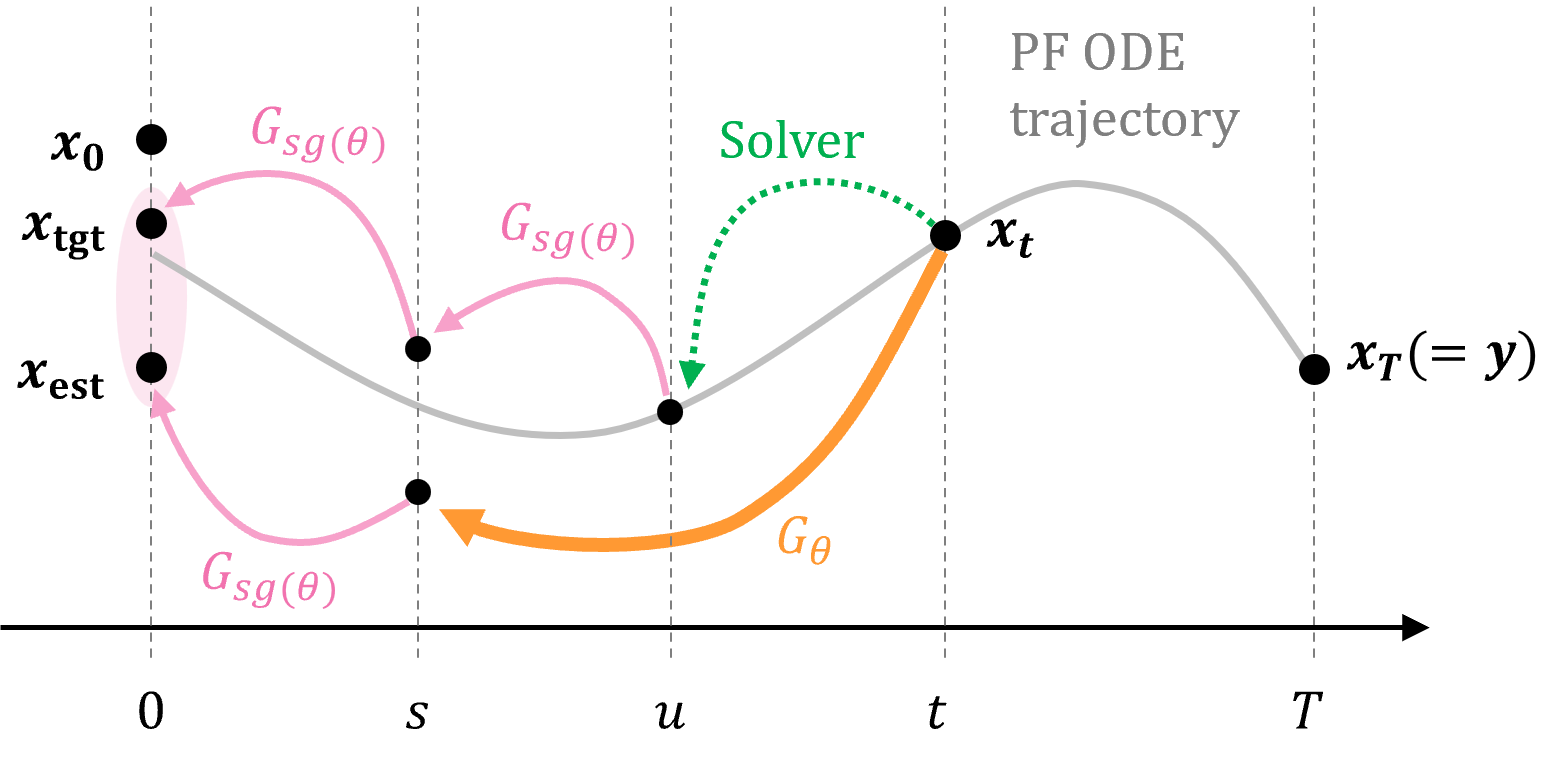}
    \caption{SBCTM training overview}
    \label{fig:framework}
\end{figure}

We trained the SBCTM based on the CTM's~\cite{ctm} training framework to estimate the complex STFT spectrum $x_0$ of clean speech from the complex STFT spectrum $y$ of noisy speech, as illustrated in Fig.~\ref{fig:framework}. To adapt to the SB~\cite{sb}, we introduced $y$ into $G_\theta$, defined as follows:
\begin{equation}
    G_\theta(x_t, y, t, s) := \frac{s}{t}x_t + (1-\frac{s}{t})F_\theta(x_t, y, t, s), \label{eq:gth}
\end{equation}
where $F_\theta$ denotes the neural network model used to estimate the process state. We define $x_\mathrm{tgt}$, $x_\mathrm{est}$, and $x_\mathrm{self}$ as follows:
\begin{align}    
    x_\mathrm{tgt} &:= G_\mathrm{sg(\theta)}(G_\mathrm{sg(\theta)}(\mathrm{Solver}(x_t, y, t, u; \phi), y, u, s), y, s, 0), \label{eq:xtgt} \\
    x_\mathrm{est} &:= G_\mathrm{sg(\theta)}(G_\theta(x_t, y, t, s), y, s, 0), \label{eq:xest} \\
    x_\mathrm{self} &:= F_\theta(x_t, y, t, t).
\end{align}
\vspace{-20pt}
\vspace{-3pt}
We introduce a novel auxiliary loss in addition to existing auxiliary losses, which comprise a time-domain loss term (TD loss) and a perceptual loss term (PESQ loss), referring to~\cite{sgmse-sb} for better performance. Furthermore, we employ $\ell_2$-norm for the original CTM's $d_\mathrm{feat}$ in Eq.~\eqref{eq:org_ctmloss}. Therefore, we modify the losses from the original CTM as follows:
\begin{align}
    \mathcal{L}_\mathrm{CTM}^\mathrm{SB} &:= \|x_\mathrm{tgt} - x_\mathrm{est}\|_2^2 + \lambda_\mathrm{td}\|\underline{x_\mathrm{tgt}} - \underline{x_\mathrm{est}}\|_1 + \lambda_\mathrm{p}\mathcal{L}_\mathrm{PESQ}(\underline{x_\mathrm{tgt}}, \underline{x_\mathrm{est}}), \label{eq:ctmloss} \\
    \mathcal{L}_\mathrm{DSM}^\mathrm{SB} &:= \|x_0 - x_\mathrm{self}\|_2^2 + \lambda_\mathrm{td}\|\underline{x_0} - \underline{x_\mathrm{self}}\|_1 + \lambda_\mathrm{p}\mathcal{L}_\mathrm{PESQ}(\underline{x_0}, \underline{x_\mathrm{self}}), \label{eq:dsmloss}
\end{align}
where $\mathcal{L}_\mathrm{PESQ}(\cdot,\cdot)$ represents a perceptual loss function~\cite{pesqloss} utilizing a differentiable version of the PESQ metric, and $\lambda_\mathrm{td}$ and $\lambda_\mathrm{p}$ denote the weights of each loss term. In $\mathcal{L}_\mathrm{PESQ}(\cdot,\cdot)$, the first input corresponds to the reference signal, and the second input corresponds to the signal being evaluated. The underlined characters represent the corresponding time-domain signals obtained using an inverse STFT. Summing Eq.~\eqref{eq:ctmloss}, and Eq.~\eqref{eq:dsmloss}, SBCTM is trained with an objective defined as follows:
\begin{equation}
    \mathcal{L} := \mathcal{L}_\mathrm{CTM}^\mathrm{SB} + \lambda_\mathrm{DSM} \mathcal{L}_\mathrm{DSM}^\mathrm{SB}. \label{eq:sbctmloss}
\end{equation}
Following CTM and SoundCTM~\cite{sctm}, we employ adaptive weighting with $\lambda_\mathrm{DSM} = \frac{\|\nabla_{\theta_L}\mathcal{L}_\mathrm{CTM}^\mathrm{SB}\|_2^2}{\|\nabla_{\theta_L}\mathcal{L}_\mathrm{DSM}^\mathrm{SB}\|_2^2}$, where $\theta_L$ denotes the last layer of the student's network.

\subsection{Inference}

The inference procedure for SBCTM starts by initializing $x_{t_N}$ with observation $y$. Then, for each step $n = N, N-1,..., 1,$ the model estimates the state at time $t_{n-1}$ by applying the transformation $G_\theta$ to $x_{t_n}$, conditioned on $y, t_n,$ and $t_{n-1},$ that is,
\begin{equation}
    x_{t_{n-1}} \leftarrow G_\theta(x_{t_n}, y, t_n, t_{n-1}).
\end{equation}
After iterating all the steps, the final output $x_{t_0}$ is returned as an estimation of $x_0$.

\section{EXPERIMENTAL SETUP} \label{sec:exp}

\subsection{Models settings} \label{ssec:model_settings}

We employed the NCSN++ architecture~\cite{ncsnpp} proposed in~\cite{sgmse-sb} as the network model $F_\theta$ in Eq.~\eqref{eq:gth}. We added a new time condition $s$ as the embedding to the model ($s$-embedding) as follows~\cite{sctm}: For the input representation, we used complex STFT spectra. We employed a periodic Hann window with $510$ samples as the window length of $128$ samples. We employed the SB with a variance exploding diffusion coefficient (SB-VE) with the recommended hyperparameters from a previous study~\cite{sb-se}.

For the teacher model, we selected the M7 model from a group of pretrained models proposed in~\cite{sgmse-sb}. We initialized the student model's parameters with the teacher model's parameters, except for the student model's $s$-embedding, following~\cite{sctm}. With respect to $\mathrm{Solver}(x_t, y, t, u;\phi)$ in Eq.~\eqref{eq:xtgt}, we used the ODE sampler proposed in~\cite{sgmse-sb} (see Equation (20) in the paper).
We used $\mathrm{sg(\theta)}$ for the inference instead of $\theta$, following the literature~\cite{sgmse}.

\subsection{Training details}

We followed the method of~\cite{sctm} for timestep sampling. To calculate the objective of SBCTM, we selected $t$ and $s$ from the $N$-discretized time steps defined as follows: $({\sigma_\mathrm{max}}^{\frac{1}{\rho}}+\frac{i}{N-1}({\sigma_\mathrm{min}}^{\frac{1}{\rho}}-{\sigma_\mathrm{max}}^{\frac{1}{\rho}})^\rho$, where $i$ denotes the index of discretized time steps.

Throughout the experiments, we used $T = 1.0$, $N = 40$, $\mu = 0.999$, $\sigma_\mathrm{min} = 3.0\times10^{-2}$, $\sigma_\mathrm{max} = 1.0$, $\rho = 7$, and the RAdam optimizer~\cite{radam} with a learning rate of $8.0\times10^{-5}$. Regarding the weights for the auxiliary loss, we set $1.0\times10^{-3}$ for $\lambda_\mathrm{td}$ and $5.0\times10^{-4}$ for $\lambda_\mathrm{p}$. The experimental results with variations in $\lambda_\mathrm{p}$ are presented in Section~\ref{subsec:wbalance}.

We used VoiceBank+Demand (VB-DMD)~\cite{vbdmd}, which is commonly employed as a benchmark for SE tasks, as the training dataset. The VB-DMD includes 28 speakers’ voices and has approximately 9.5 h of training data. We used 9 h of data for training and the remaining data for validation. All audio samples were downsampled to 16 kHz. We trained the model for 100 epochs (69.3 K training steps), except for the ablation study, with a global batch size of 16, using 8 $\times$ NVIDIA A100 GPUs, each of which had 40 GB memory.

\subsection{Evaluation dataset and metrics}

To evaluate the models, we used the official test data of the VB-DMD, which contained approximately 0.5 h of data. We used PESQ with an extended bandwidth~\cite{pesq} as an intrusive SE metric. We employed the scale-invariant signal-to-noise ratio (SI-SDR) to observe the similarity to the waveform of clean speech. Furthermore, we used ESTOI~\cite{estoi} as an instrumental measure of speech intelligibility. As a non-intrusive metric, we used DNSMOS~\cite{dnsmos} based on ITU-T P.808~\cite{p808}, which employed a neural network trained on human ratings. For all metrics, it holds that the higher, the better.

We evaluated each model based on a comprehensive judgment in terms of each metric while prioritizing PESQ, as PESQ not only has a higher correlation with our auditory perception compared with other metrics such as SI-SDR\cite{se-odg}, but also, to the best of our knowledge, it is the most commonly used metric in various studies on SE~\cite{demucs, dptfsnet, cmgan, sgmse, mpsenet, sgmse-sb, semamba}.

\section{RESULTS} \label{sec:res}

\subsection{Comparison with teacher model and other consistency-based models} \label{ssec:cmp_with_teacher}

We first compared our SBCTM with the teacher model in terms of the effectiveness of the CTM training~\cite{ctm}. As explained in Section~\ref{ssec:model_settings}, we selected the pretrained model (M7) proposed in SB-PESQ~\cite{sgmse-sb} as the teacher model. The teacher model was evaluated to determine the metrics for each NFE. Regarding the student models, we introduced Schr\"odinger bridge consistency models (SBCM) as an additional reference for comparison. We applied consistency training~\cite{cm} to train the SBCM by employing the same teacher model as the SBCTM. Furthermore, we included the SE-Bridge~\cite{seb} in our comparison as a baseline, which introduces consistency training into the Brownian bridge.

Table~\ref{tab:result} summarizes the results for the SE task using the VB-DMD dataset~\cite{vbdmd} for both the teacher and student models. PESQ, SI-SDR, and DNSMOS of the SBCTM outperformed those of the teacher model when $\mathrm{NFE}=1$. Furthermore, the performance of the SBCTM improved as the NFE increased. Similar to what was reported in~\cite{ctm} and~\cite{sctm}, this effect was also observed in the SBCTM. Regarding the PESQ in Table~\ref{tab:result}, the SBCTM achieved a higher score than the teacher model for all NFEs. This is attributed to the auxiliary loss in Eqs.~\eqref{eq:ctmloss} and~\eqref{eq:dsmloss}, including the PESQ loss added to the original loss~\cite{cm}. Details are presented in Section~\ref{ssec:ablation}.
Comparing the model with other consistency-based models, the SE-Bridge achieved a high SI-SDR even with $\mathrm{NFE}=1$, whereas the SBCTM demonstrated superior performance in terms of PESQ. The SBCM's SI-SDR outperformed that of the teacher model when $\mathrm{NFE}=1$, but most scores exhibited a decreasing trend as NFE increased. However, this can be attributed to our simple training strategies.

Regarding the practical inference speed, we calculated the RTF of both the SB-PESQ ($\mathrm{NFE}=16$) and the SBCTM ($\mathrm{NFE}=1$), which exhibited comparable performance in terms of PESQ, ESTOI, and DNSMOS metrics. The RTF was measured using a Linux PC equipped with an NVIDIA GeForce RTX4080 (16 GB) GPU. We evaluated the RTF as the average over 10 runs using 10-s audio inputs with a batch size of 1. The results demonstrated that the SB-PESQ achieved an RTF of 0.713, whereas the SBCTM achieved an RTF of 0.045. This indicates that SBCTM achieves approximately $16\times$ improvement in the inference speed compared with SB-PESQ.
\begin{table}[ht]
    \centering
    \caption{SE performance on the VB-DMD. Each value of the metrics represents mean score, where higher values indicate better quality. SB-PESQ$^{1}$ refers to the originally proposed model (M7) reported in~\cite{sgmse-sb}; the NFE is not specified. SB-PESQ$^{2}$ denotes that the model is evaluated by us. SE-Bridge refers to the originally proposed model reported in~\cite{seb}; DNSMOS is not reported.}
    \label{tab:result}
    \begin{threeparttable}
        \begin{tabularx}{\columnwidth}{X c c c c c}
            \toprule
            Model & NFE & PESQ & SI-SDR & ESTOI & DNSMOS \\
            \midrule
            Input & - & $1.96$ & $8.4$ & $0.79$ & $3.08$ \\
            \midrule
            SB-PESQ\tnote{1} & - & $3.50$ & $14.1$ & $\textbf{0.87}$ & $\textbf{3.55}$ \\
            SB-PESQ\tnote{2} & 1 & $3.45$ & $10.6$ & $\textbf{0.87}$ & $3.51$ \\
            & 2 & $3.52$ & $12.0$ & $\textbf{0.87}$ & $3.52$ \\
            & 4 & $3.55$ & $13.0$ & $\textbf{0.87}$ & $3.54$ \\
            & 8 & $3.56$ & $13.2$ & $\textbf{0.87}$ & $3.54$ \\
            & 16 & $3.56$ & $13.2$ & $\textbf{0.87}$ & $\textbf{3.55}$ \\
            \midrule
            SE-Bridge & 1 & 2.97 & \textbf{19.9} & \textbf{0.87} & - \\
            SBCM & 1 & $3.00$ & $14.0$ & $0.84$ & $3.47$ \\
            &  2 & $1.71$ & $ 9.2$ & $0.74$ & $2.89$ \\
            &  4 & $1.87$ & $ 5.6$ & $0.73$ & $3.04$ \\
            &  8 & $2.01$ & $ 3.3$ & $0.72$ & $3.16$ \\
            & 16 & $1.97$ & $ 1.6$ & $0.71$ & $3.16$ \\
            \rowcolor{blue!10} \textbf{SBCTM (Ours)} & 1 & $3.56$ & $12.2$ & $\textbf{0.87}$ & $\textbf{3.55}$ \\
            \rowcolor{blue!10} & 2 & $3.54$ & $13.2$ & $\textbf{0.87}$ & $3.54$ \\
            \rowcolor{blue!10} & 4 & $3.57$ & $12.8$ & $\textbf{0.87}$ & $3.54$ \\
            \rowcolor{blue!10} & 8 & $3.57$ & $12.7$ & $\textbf{0.87}$ & $3.53$ \\
            \rowcolor{blue!10} & 16 & $\textbf{3.58}$ & $12.7$ & $\textbf{0.87}$ & $3.53$ \\
            \bottomrule
        \end{tabularx}
    \end{threeparttable}
\end{table}

\subsection{Ablation study on effect of auxiliary loss} \label{ssec:ablation}

Table~\ref{tab:ablation} summarizes the results of an ablation study on the auxiliary loss in Eqs.~\eqref{eq:ctmloss} and ~\eqref{eq:dsmloss}. We trained all models in Table~\ref{tab:ablation} on the VB-DMD training dataset with 20 epochs and observed the inference results when $\mathrm{NFE}=1$.

Observing c) in the results, the auxiliary loss significantly contributes to PESQ, whereas the SI-SDR decreases.
a) illustrates that the auxiliary CTM loss contributes to the SI-SDR;
b) shows that the auxiliary loss of the DSM loss contributes significantly to PESQ while decreasing the SI-SDR.
Furthermore, the results for d) and e) show that the TD loss contributes to the SI-SDR, whereas the PESQ loss contributes to PESQ.
From these results, it can be observed that all the elements of the auxiliary loss are necessary to achieve the highest SI-SDR while ensuring the highest PESQ.
\begin{table}[ht]
    \centering
    \caption{Ablation study on auxiliary loss. Values indicate mean of PESQ and SI-SDR, where higher values indicate better quality. ``w/o'' means without.}
    \label{tab:ablation}
    \begin{tabular}{l c c}
        \toprule
        Method & PESQ & SI-SDR \\
        \midrule
        Objective $\mathcal{L}$ (in Eq.~\eqref{eq:sbctmloss}) & \textbf{3.57} & 12.6 \\
        \midrule
        a) w/o auxiliary loss of $\mathcal{L}_\mathrm{CTM}^\mathrm{SB}$ & \textbf{3.57} & 12.2 \\
        b) w/o auxiliary loss of $\mathcal{L}_\mathrm{DSM}^\mathrm{SB}$ & 3.22 & 18.3 \\
        c) w/o auxiliary loss of $\mathcal{L}_\mathrm{CTM}^\mathrm{SB}$ and $\mathcal{L}_\mathrm{DSM}^\mathrm{SB}$ & 3.13 & \textbf{18.9} \\
        \midrule
        d) w/o TD loss of $\mathcal{L}_\mathrm{CTM}^\mathrm{SB}$ and $\mathcal{L}_\mathrm{DSM}^\mathrm{SB}$ & \textbf{3.57} & 12.0 \\
        e) w/o PESQ loss of $\mathcal{L}_\mathrm{CTM}^\mathrm{SB}$ and $\mathcal{L}_\mathrm{DSM}^\mathrm{SB}$ & 3.19 & 17.3 \\
        \bottomrule
    \end{tabular}
\end{table}

\subsection{Comparison of weight balance regarding auxiliary loss} \label{subsec:wbalance}

\begin{figure}[ht]
    \centering
    \includegraphics[width=0.95\linewidth]{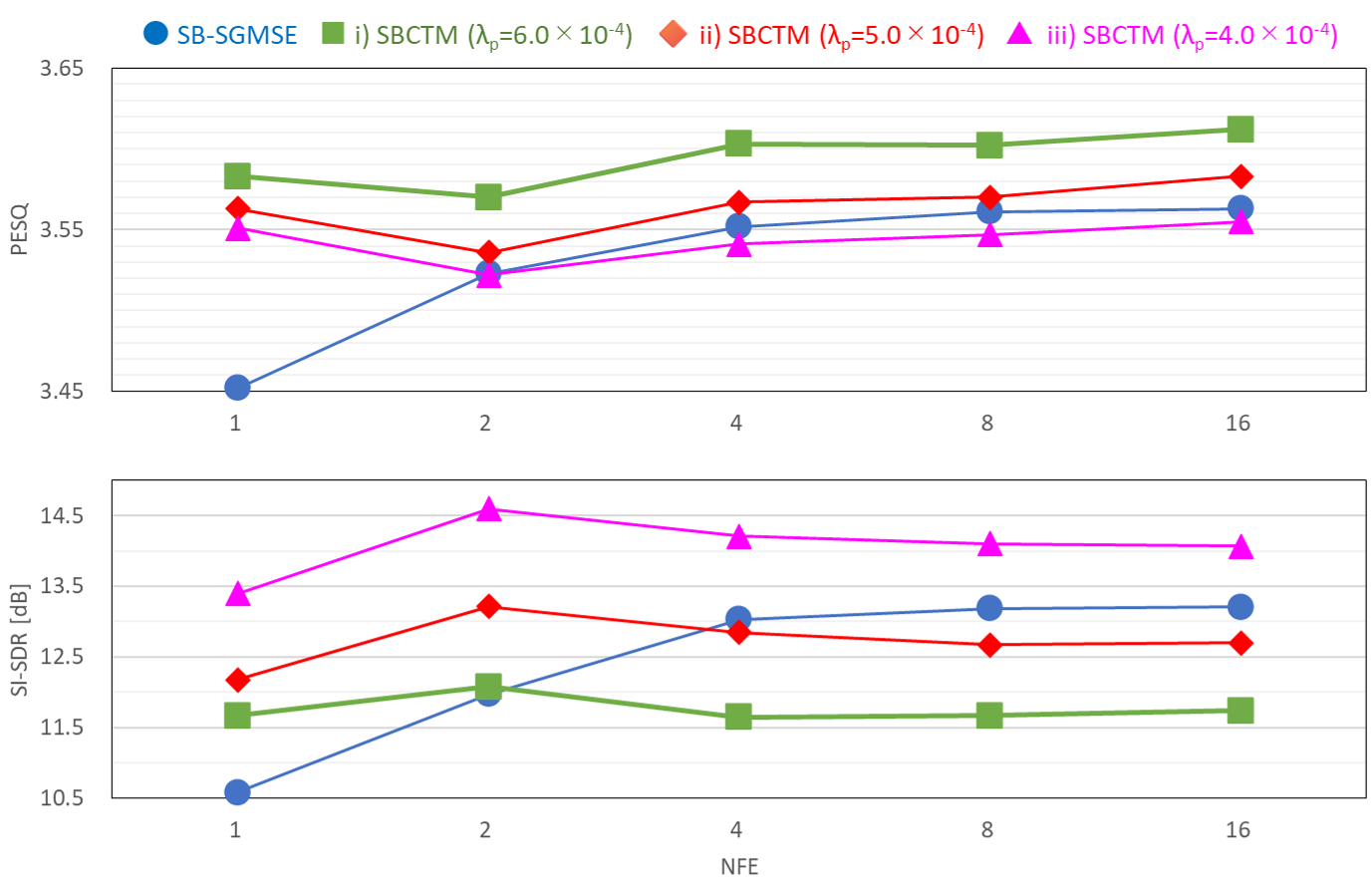}
    \caption{Comparison between SB-PESQ and variations of SBCTM with changes in $\lambda_\mathrm{p}$. Upper chart shows mean of PESQ, lower chart shows mean of SI-SDR.}
    \label{fig:lambda_pesq}
\end{figure}

We observed the effect of varying the weight balance of the auxiliary loss on the results. Fig.~\ref{fig:lambda_pesq} illustrates PESQ and SI-SDR for each NFE, including the results of the SB-PESQ as the teacher model, and the SBCTM when the value of $\lambda_\mathrm{p}$ in Eqs. ~\cref{eq:ctmloss,eq:dsmloss} were changed to i) $6.0\times10^{-4}$, ii) $5.0\times10^{-4}$, and iii) $4.0\times10^{-4}$.

PESQ can be observed to be positively correlated with $\lambda_\mathrm{p}$, whereas SI-SDR is negatively correlated. From the result of Fig.~\ref{fig:lambda_pesq}, the quality balance between PESQ and SI-SDR can be controlled by changing $\lambda_\mathrm{p}$. This enables task-specific optimization. For instance, PESQ is important for telephony, where perceptual quality is crucial for user experience. In contrast, higher SI-SDR benefits downstream tasks such as speech separation, where signal fidelity plays a more significant role. This flexibility in adjusting $\lambda_\mathrm{p}$ for task-specific objectives extends the utility of SBCTM across a wider range of applications.

\subsection{Comparison with other speech enhancement models}

Table~\ref{tab:cmp_other_models} summarizes the results of the comparison between SBCTM and other SE models in terms of the performance of PESQ using the VB-DMD dataset and the number of model parameters for each model.

Regarding performance metrics, PESQ is only commonly reported in the original studies of each model. For SEMamba~\cite{semamba} and SEMamba(+PCS)~\cite{semamba}, the RTF was measured using the method described in Section~\ref{ssec:cmp_with_teacher} and both yielded the same result of 0.059.

From the perspective of RTF, comparing the value of 0.045 for the SBCTM ($\mathrm{NFE}=1$) described in Section~\ref{ssec:cmp_with_teacher} with the value of 0.059 for SEMamba, the SBCTM ($\mathrm{NFE}=1$) demonstrates sufficiently fast inference despite its large number of parameters. In terms of quality, its performance is comparable to that of state-of-the-art methods.
\begin{table}[ht]
    \centering
    \caption{SE performance of other conventional models on VB-DMD. Values indicate mean of PESQ, where higher values indicate better quality. \# of params represents number of trainable parameters in million. SB-PESQ refers to the originally proposed model (M7) reported in~\cite{sgmse-sb}}
    \label{tab:cmp_other_models}
    \begin{tabular}{l c c}
        \toprule
        Model & \# of params [M] & PESQ \\
        \midrule
        DEMUCS~\cite{demucs} & 33.5 & 3.07 \\
        MP-SENet~\cite{mpsenet} & 2.1 & 3.50 \\
        DPT-FSNet~\cite{dptfsnet} & 0.88 & 3.33 \\
        CMGAN~\cite{cmgan} & 1.8 & 3.41 \\
        SGMSE+~\cite{sgmse} & 65.6 & 2.93 \\
        SB-PESQ~\cite{sgmse-sb} & 65.6 & 3.50 \\
        SEMamba~\cite{semamba} & 2.3 & 3.55 \\
        SEMamba(+PCS)~\cite{semamba} & 2.3 & \textbf{3.69} \\
        \midrule
        \textbf{SBCTM ($\mathrm{NFE}=1$)} & 66.0 & 3.56 \\
        \bottomrule
    \end{tabular}
\end{table}

\section{CONCLUSION} \label{sec:con}

We propose SBCTM, which achieves high-quality one-step inference while maintaining a favorable trade-off between quality and inference speed in SE.
The proposed auxiliary loss contributes to improving performance. Experimental results demonstrate that SBCTM achieves approximately $16\times$ improvement in RTF compared with the conventional SB for SE. The proposed approach enables efficient and practical SE, particularly in latency-sensitive applications. Furthermore, SBCTM can potentially be extended to other tasks such as dereverberation and declipping. By achieving fast and high-quality inference with controllable trade-offs, our method offers a new direction for SE research.

\clearpage
\bibliographystyle{IEEEtran}
\bibliography{refs25}

\end{document}